\documentclass[twocolumn,twoside,slac_two]{revtex4}
\usepackage{graphicx}
\usepackage{fancyhdr}
\begin{document}

\title{$\phi_2/\alpha$}

%

\author{C.C.Wang}
\affiliation{Department of Physics, National Taiwan University, 
No.1, Sec.4, Roosevelt Rd., Taipei, Taiwan, 106}

\begin{abstract}
We report the recent $\phi_2/\alpha$ results from Belle with KEKB accelerator
and Babar with PEP-II accelerator. The analysis of $B\to \pi\pi$, $B\to \rho\rho$
and $B\to \rho\pi$ are included in this report.
These $b\to u\bar{u}d$ decay modes are related to the CKM angle $\phi_2/\alpha$ 
and the method of $\phi_2/\alpha$ extraction of corresponding decays are also 
included.
After combining all the decay modes, the constraint of $\phi_2/\alpha$ is 
$(100.2^{+15.0}_{-8.8})^\circ$.
\end{abstract}

\maketitle

\thispagestyle{fancy}


\section{Introduction}
In the Standard Model (SM), $CP$ violating effects 
in the B meson system can be parameterized in terms of 
three Cabibbo-Kobayashi-Maskawa (CKM)~\cite{bib:KM} phase angles 
$\phi_1$, $\phi_2$ and $\phi_3$ (which can be also written 
as $\beta$, $\alpha$ and 
$\gamma$, respectively).
The angle 
$\phi_2/\alpha$ can be extracted via the $b\to u\bar{u}d$.
Approaches for $\phi_2/\alpha$ extraction
from the isospin analysis, $B\to \pi\pi$ and 
$B\to \rho\rho$, and the time-dependent Dalitz analysis, $B\to \rho\pi$ are reported in
this document. 

The experimental measurements are from B-factories, Belle detector with KEKB 
accelerator and Babar detector with PEP-II accelerator. These two detectors
are general purpose detectors with energy-asymmetric $e^+e^-$ accelerators.
The Belle detector consists of a silicon vertex detector
(SVD), a central drift chamber (CDC), an array of aerial threshold 
\v{C}erenkov counters (ACC), time-of-flight scintillation counters (TOF), 
and an electromagnetic calorimeter (ECL) comprised of CsI(Tl) crystals located 
inside a superconducting solenoid coil
that provides a 1.5~T magnetic field. An iron flux return located outside of
the coil is instrumented to detect $K_L^0$ mesons and identify muons. 
The Babar detector contains silicon vertex tracker (SVT), drift chamber, 
electromagnetic calorimeter, ring-imaging \v{C}erenkov detector (DIRC) and 
a 1.5~T solenoid superconducting magnet.

\section{$B\to\pi\pi$}
In the $B^0\to\pi^+\pi^-$ decay, the time dependent rate is described by
\begin{eqnarray}
\nonumber
\mathcal{P}^q(\Delta t) & =  &{e^{-|\Delta t|/\tau_{B^0}}\over 4\tau_{B^0}}
[1+q\cdot \{\mathcal{S}\rm{sin}(\Delta m_d\Delta t)+\\
& & \mathcal{A}\rm{cos}(\Delta m_d\Delta t)\}]. 
\label{eqn:dt_pdf}
\end{eqnarray}
The notation $\mathcal{A}$ is used by Belle collaboration which is
the same as $-\mathcal{C}$ used by Babar collaboration.

\begin{figure}[h]
\centering
\includegraphics[width=35mm]{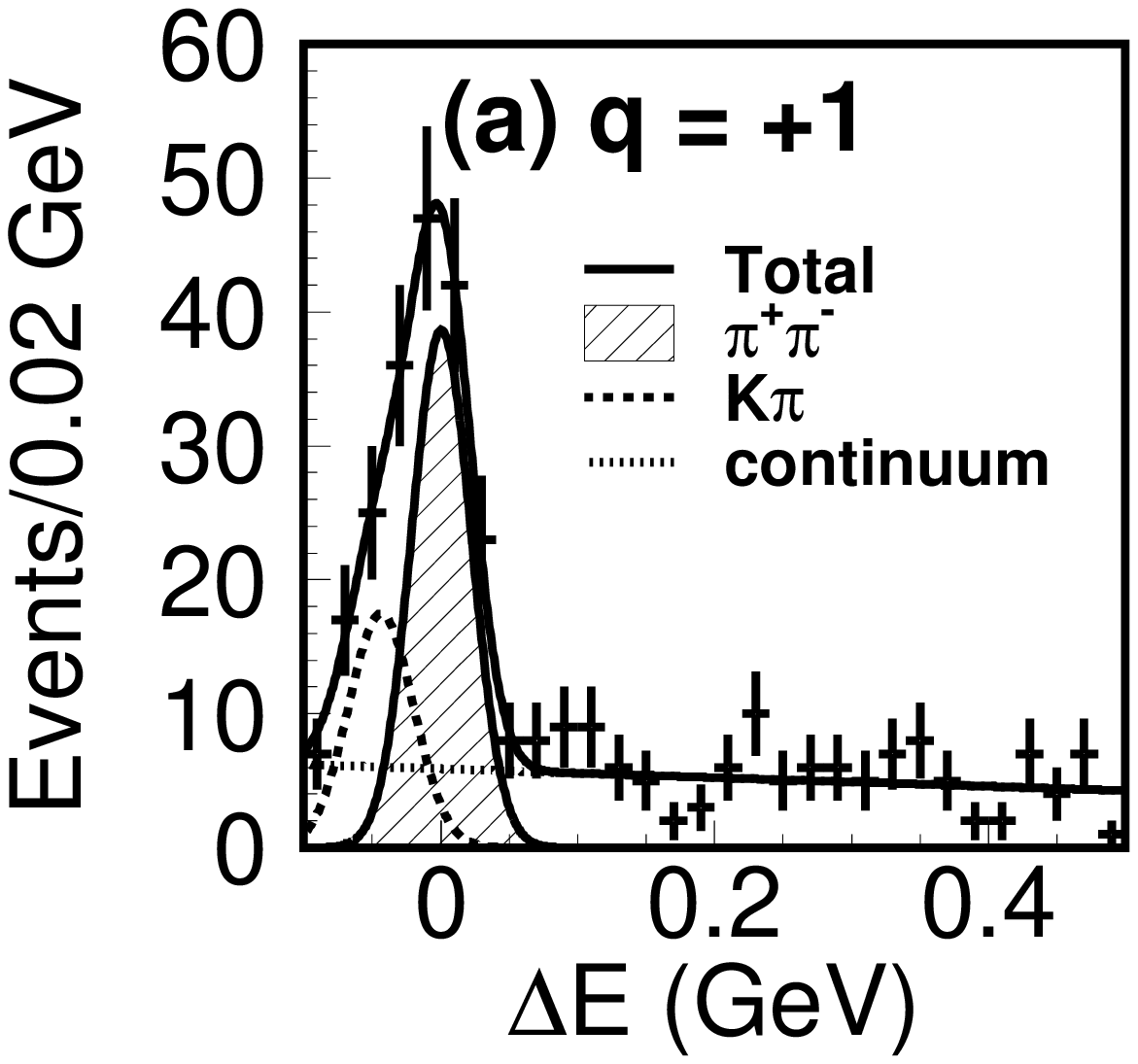}
\includegraphics[width=35mm]{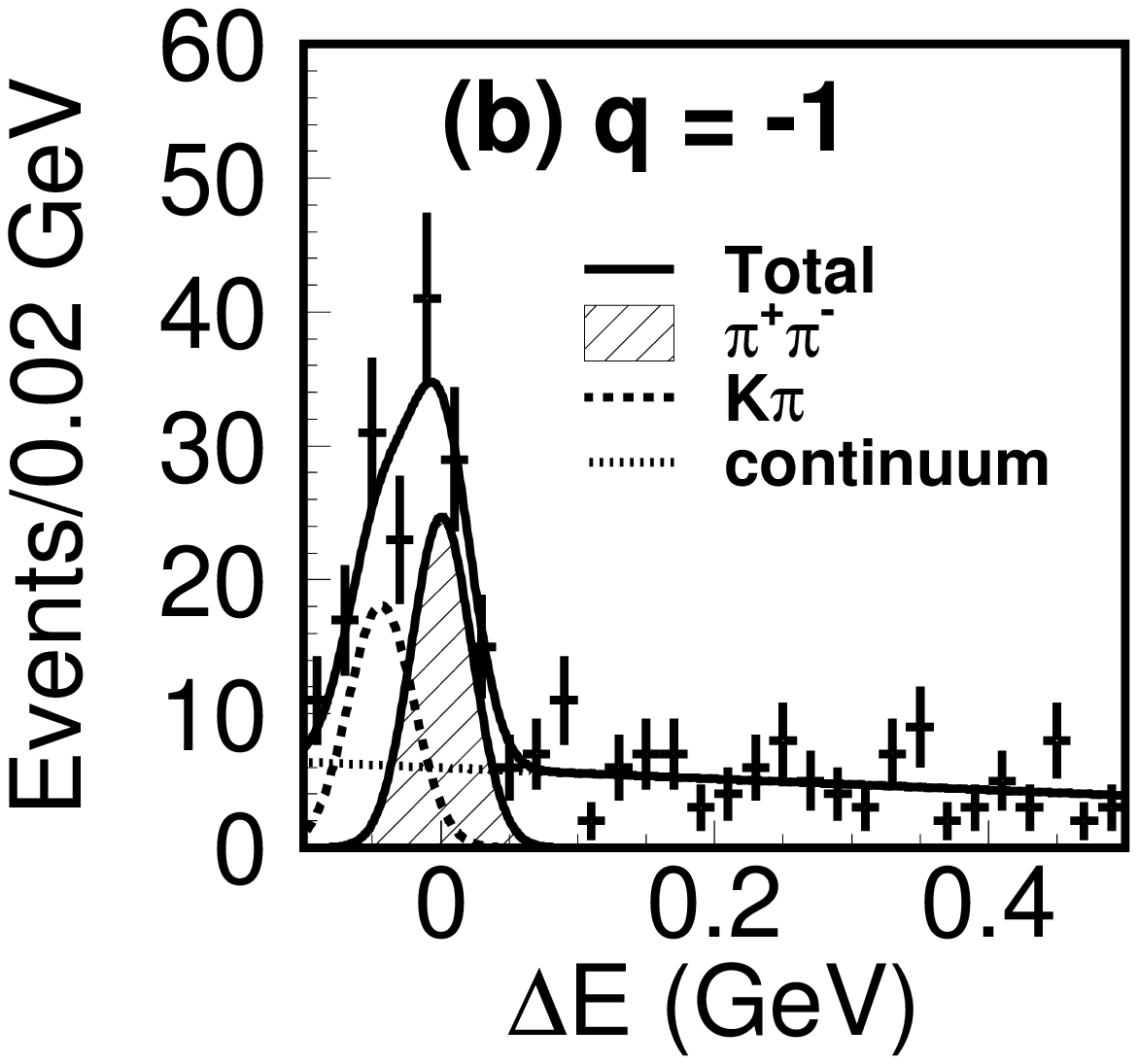}
\caption{$\Delta E$ distributions in the $M_{bc}$ signal region 
(5.271 GeV/c$^2$ $<$ $M_{bc}$ $<$ 5.287 GeV/c$^2$) for
$B^0\to \pi^+\pi^-$ candidates with $\mathcal{LR}>0.86$ for (a) 
$q = +1$ and (b) $q = -1$ from Belle. } 
\label{pipi_de_belle}
\end{figure}

\begin{figure}[h]
\centering
\includegraphics[width=60mm]{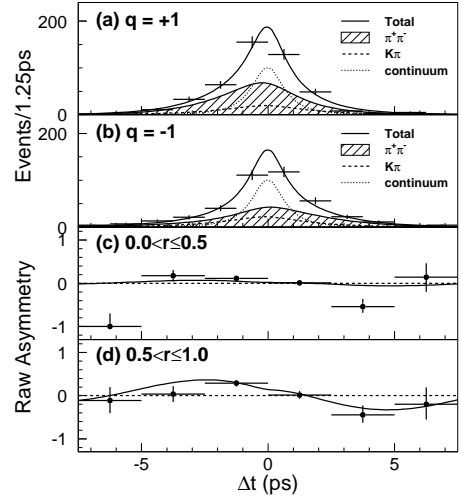}
\caption{The Belle $B^0\to \pi^+\pi^-$ $\Delta t$ distributions 
for the candidates with $\mathcal{LR}>0.86$ in the signal region 
(5.271 GeV/c$^2$ $<$ $M_{bc}$ $<$ 5.287 GeV/c$^2$  and 
$|\Delta E|$ $<$ 0.064GeV). (a) $q = +1$ and (b) $q = -1$.
Raw asymmetry, $\mathcal{A}_{cp}$, in each $\Delta t$ bin with
(c) 0$<$r$\le$0.5 and (d)0.5$<$r$\le$1.0. The solid lines shows the
result of the unbinned maximum likelihood fit. } 
\label{pipi_dt_belle}
\end{figure}

The $B^0\to \pi^+\pi^-$ analysis uses 253~fb$^{-1}$ data collected 
by Belle~\cite{bib:belle_pipi}. Signal candidates are reconstructed
by opposite charged tracks which are identified as pions and  
the pion 
identification is based on the combined information from the ACC and the CDC
$dE/dx$ measurements. $B$ meson candidates are selected by using the 
energy difference $\Delta E\equiv E^*_B - E^*_{beam}$ and the beam-energy 
constrained mass $M_{bc}\equiv \sqrt{(E^*_{beam})^2-(p^*_B)^2}$, where 
$E^*_{beam}$ is the CMS beam-energy, and $E^*_B$ and $p^*_B$ are the CMS 
energy and momentum of the $B$ candidate, respectively. 
The flavor of accompanying $B$ meson
is identified from the inclusive properties of particles which are not 
used for $B^0\to \pi^+\pi^-$ reconstruction. To suppress the continuum 
background ($e^+e^-\to q\bar{q}$; q=u,d,s,c), the likelihood ratio 
($\mathcal{LR}$) of the event topology based on signal MC and sideband data 
is used for continuum suppression. The likelihood ratio is optimized separately 
for each flavor
tagging quality region. The tagging quality is monitored by variable r. 
After applying all above requirement and vertex
reconstruction algorithm, 2820 signal candidates containing 
$666\pm 43$ $\pi^+\pi^-$ signal events (1486 $B^0$ tags and 1334 
$\overline{B^0}$ tags) are obtained. 
The $CP$ violation parameters $\mathcal{S}$
and $\mathcal{A}$ are determined from the unbinned maximum likelihood fit 
to the proper-time difference $\Delta t$ distribution. They obtained
$\mathcal{S} = -0.67\pm0.16\rm{(stat)}\pm0.06\rm{(syst)}$ and 
$\mathcal{A} = +0.56\pm0.12\rm{(stat)}\pm0.06\rm{(syst)}$. The 
correlation between $\mathcal{S}$ and $\mathcal{A}$ 
is $+0.09$. Figure~\ref{pipi_de_belle} and Figure~\ref{pipi_dt_belle} 
show the result from Belle.

\begin{figure}[h]
\centering
\includegraphics[width=70mm]{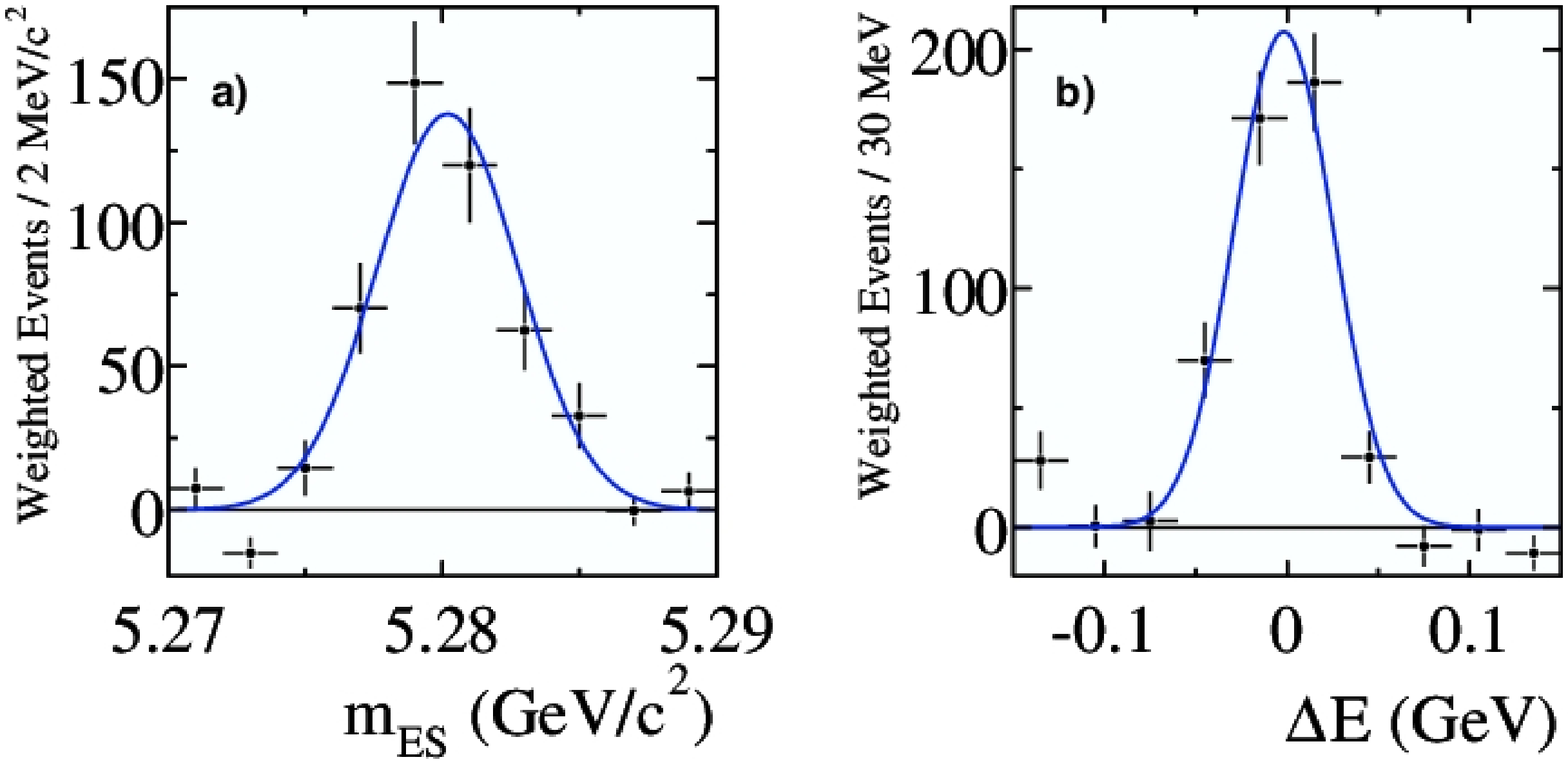}
\caption{Babar's (a) $m_{ES}$ and (b) $\Delta E$ distributions for 
$B^0\to\pi^+\pi^-$. Solid curves represent the corresponding PDFs 
used in the fit.} 
\label{pipi_de_babar}
\end{figure}

\begin{figure}[h]
\centering
\includegraphics[width=60mm]{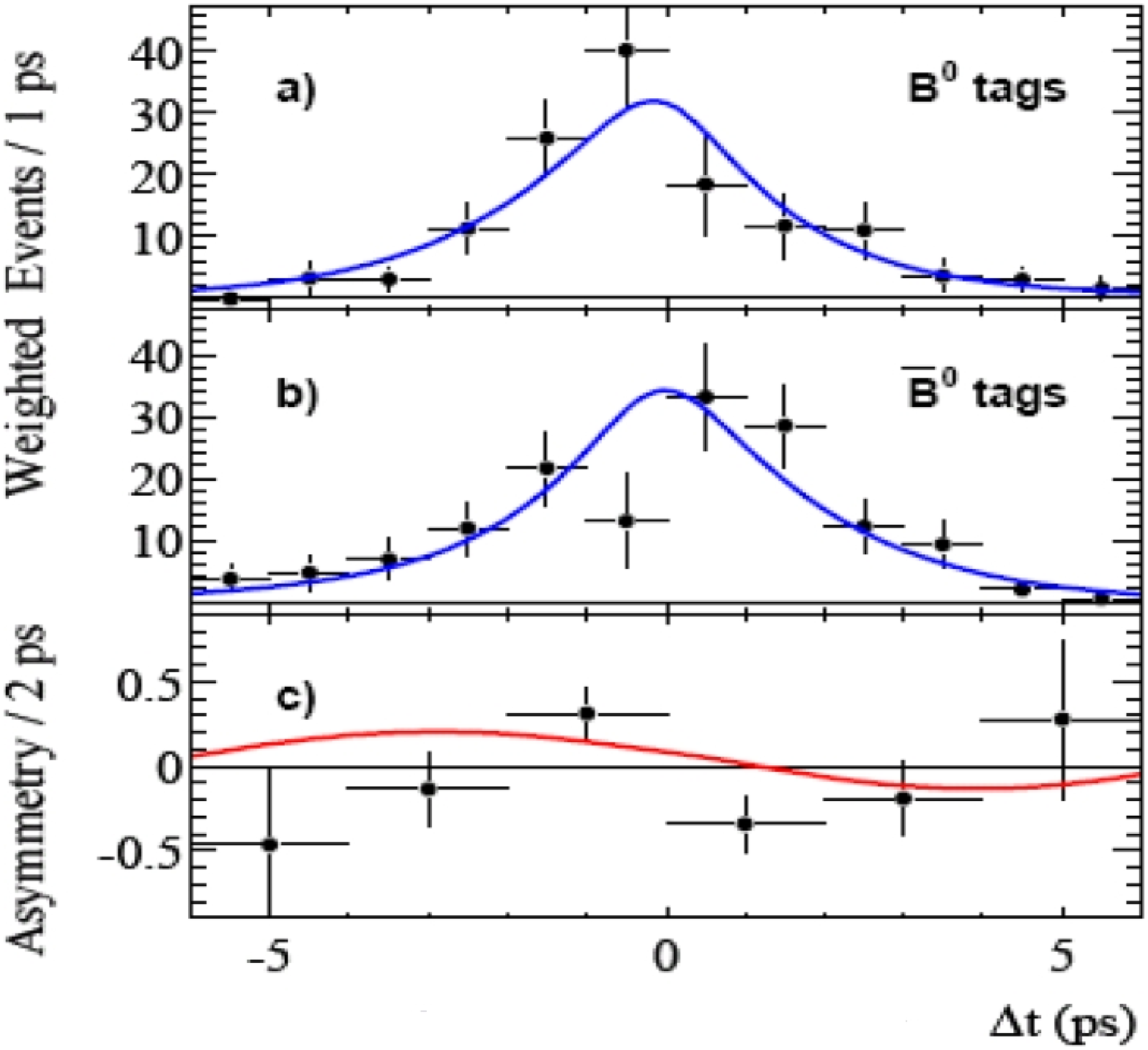}
\caption{Babar's $\Delta t$ distributions for $B^0\to\pi^+\pi^-$.
(c) raw asymmetry for signal events in each $\Delta t$ bin.} 
\label{pipi_dt_babar}
\end{figure}

The Babar's $B^0\to \pi^+\pi^-$ analysis~\cite{bib:babar_pipi} is
based on 211~fb$^{-1}$ data. Signal candidates are reconstructed from 
opposite charged tracks with associated \v{C}erenkov angles ($\theta _c$). 
The particle identification is primarily accomplished by including $\theta_c$
in the maximum-likelihood fit. The kinematic variables $\Delta E$ and 
$M_{ES} = \sqrt{(s/2+\bf{p_i}\cdot \bf{p_B})^2/E_i^2 - \bf{p_B}^2}$ are used to identify the
signal decays. Here $\bf{p_B}$ is the $B$ momentum and ($E_i$, $\bf {p_i}$) 
is the four-momentum of $e^+e^-$ initial state in lab frame. The background
suppression is based on the angle of sphericity axes and Fisher discriminant 
$\mathcal{F}$ formed from the momentum flow relative to $\pi^+\pi^-$ thrust axis.
Unbinned extended maximum-likelihood fit is used to extract $CP$ parameters and the 
likelihood function includes event yield, tagging efficiency, $m_{ES}$, 
$\Delta E$, $\mathcal{F}$, $\theta_c^+$, $\theta_c^-$ and $\Delta t$. 
The fit yields
$\mathcal{S} = -0.30\pm 0.17(\rm{stat})\pm0.03(\rm{syst})$
and $\mathcal{C} = -0.09\pm 0.15(\rm{stat})\pm0.04(\rm{syst})$ from 
$467\pm33$ $B^0 \to \pi^+\pi^-$ events.
Figure~\ref{pipi_de_babar} and Figure~\ref{pipi_dt_babar} 
show the result from Babar.  

Using the model-independent isospin analysis~\cite{bib:iso_ana1,bib:iso_ana2}, 
the range of $[19^\circ ,71^\circ]$ for $\phi_2/\alpha$
is excluded by Belle with 94.5\% C.L. whereas $[29^\circ ,61^\circ]$ is excluded by
Babar with 90\% C.L.

\section{$B\to\rho\rho$}
\begin{figure}[h]
\centering
\includegraphics[width=50mm]{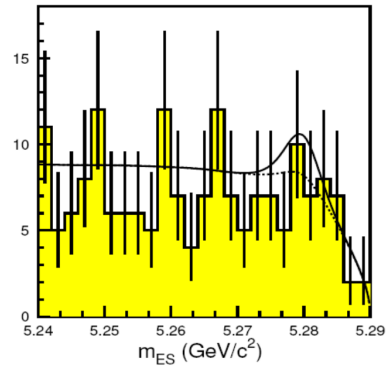}
\caption{Projection of $B^0\to\rho^0\rho^0$ $m_{ES}$ distribution obtained
at Babar.} 
\label{rho0rho0_babar}
\end{figure}

The Babar's analysis of $B^0\to\rho^0\rho^0$~\cite{bib:babar_rho0rho0}
is performed with 211~fb$^{-1}$ data. The reconstruction is made with 
four charged tracks and the particle identification is provided
by the combining information from DIRC and SVT. The continuum suppression
is performed based on the angle of thrust axis and neural network output variable
$\mathcal{E}$. Unbinned maximum likelihood fit is used to extract the 
branching fraction by combining $m_{ES}$, $\Delta E$, $\pi^+\pi^-$ 
invariant mass, $\rho$ helicity angle and flavor tagging. From the fit,
they obtain the upper limit of $1.1\times10^{-6}$ at 90\% C.L. 
The $m_{ES}$ distribution is shown in Figure~\ref{rho0rho0_babar}.

The time dependent rate of $B^0\to\rho^+\rho^-$ can be parametrized by
the same form as Eq.~\ref{eqn:dt_pdf}.  

\begin{figure}[h]
\centering
\includegraphics[width=80mm]{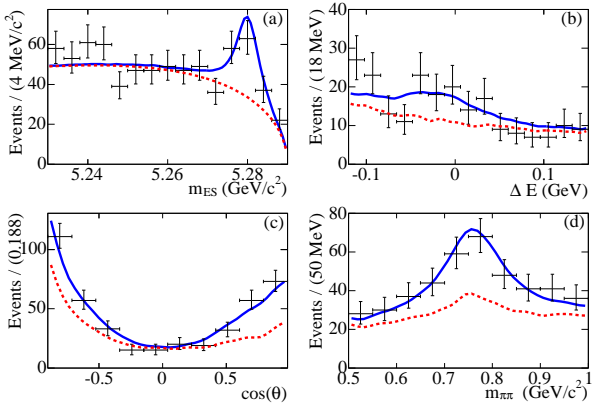}
\caption{The distribution for the high purity events for variables 
(a) $m_{ES}$, (b) $\Delta E$, (c) cosine of $\rho$ helicity angle, and
(d) $m_{\pi^\pm\pi^0}$. The dotted curves are the sum for all backgrounds 
and the solid lines are the total PDF.} 
\label{rhorho1_babar}
\end{figure}

\begin{figure}[h]
\centering
\includegraphics[width=70mm]{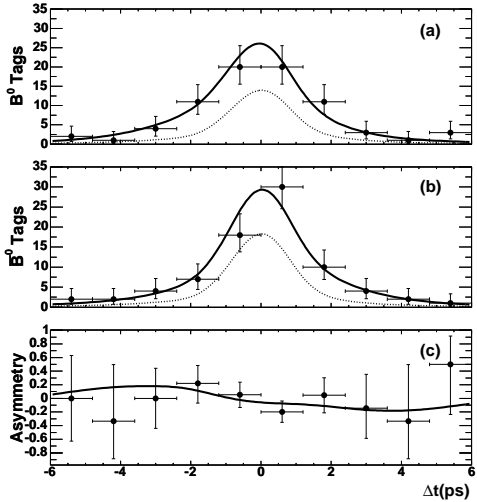}
\caption{The $\Delta t$ distribution for the signal enriched sample. (a) is $B^0$ tagged 
and (b) is $\overline{B}^0$ tagged events. (c) is the $\Delta t$ raw asymmetry.} 
\label{rhorho2_babar}
\end{figure}

The time-dependent $CP$ analysis from Babar is performed with  
211~fb$^{-1}$~\cite{bib:babar_rhorho}. Comparing to $B\to\pi^+\pi^-$
analysis, the $B^0\to\rho^+\rho^-$ need an angular analysis to extract the
fraction for longitudinal component ($f_L$). 
An extended maximum-likelihood fit provides 
$f_L=0.978\pm0.014\rm{(stat)}^{+0.021}_{-0.029}\rm{(syst)}$ and 
time-dependent $CP$ parameters, 
$\mathcal{C}_L=-0.03\pm0.18\rm{(stat)}\pm0.09\rm{(syst)}$  
and $\mathcal{S}_L=-0.33\pm0.24\rm{(stat)}^{+0.08}_{-0.14}\rm{(syst)}$.
Figure~\ref{rhorho1_babar} and Figure~\ref{rhorho2_babar} 
show the result of Babar's analysis. From the isospin analysis, Babar
obtains $\phi_2/\alpha$ between $79^\circ$ and $123^\circ$ 
with 90\% C.L. which is based on the results of $B^0\to\rho\rho$ 
mentioned above, the branching fraction of 
$B^0\to\rho^+\rho^-$~\cite{bib:babar_rhorho_iso2,bib:babar_rhorho_iso3} and the results
of $B^+\to\rho^+\rho^0$ analysis~\cite{bib:babar_rhorho_iso1, bib:belle_rhorho_iso1}.

\begin{figure}[h]
\centering
\includegraphics[width=80mm]{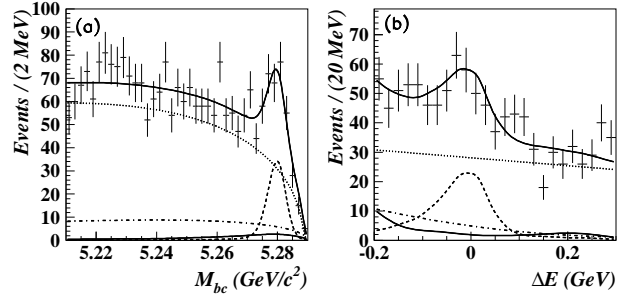}
\caption{(a) $M_{bc}$ projection in 
$-0.10\rm{GeV}<\Delta E<0.06\rm{GeV}$ region. (b) $\Delta E$
projection in $M_{bc}$ signal region.
The dashed, dotted, dot-dashed, small solid and large solid curves show
$\rho^+\rho^-+\rho\pi\pi$, $q\bar{q}$, $b\to c$, $b\to u$ and the total,
respectively.} 
\label{rhorho1_belle}
\end{figure}

\begin{figure}[h]
\centering
\includegraphics[width=80mm]{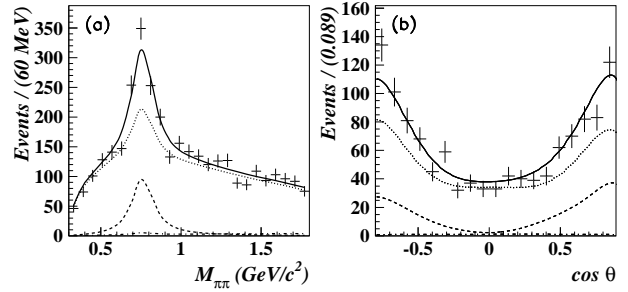}
\caption{(a) $M_{\pi^\pm \pi^0}$ projection.
(b) sum of two cosine helicity angle distribution. Both plots 
are inside $M_{bc}-\Delta E$ 
signal region and satisfy $0.62\rm{GeV/c^2}<M_{\pi\pi^0}<0.92\rm{GeV/c^2}$.
The dashed, dot-dashed, dotted and solid curves represent $\rho^+\rho^-$,
$\rho\pi\pi$, $q\bar{q}$ + $(b\to c)$ + $(b\to u)$ and the total, respectively.} 
\label{rhorho2_belle}
\end{figure}

\begin{figure}[h]
\centering
\includegraphics[width=40mm]{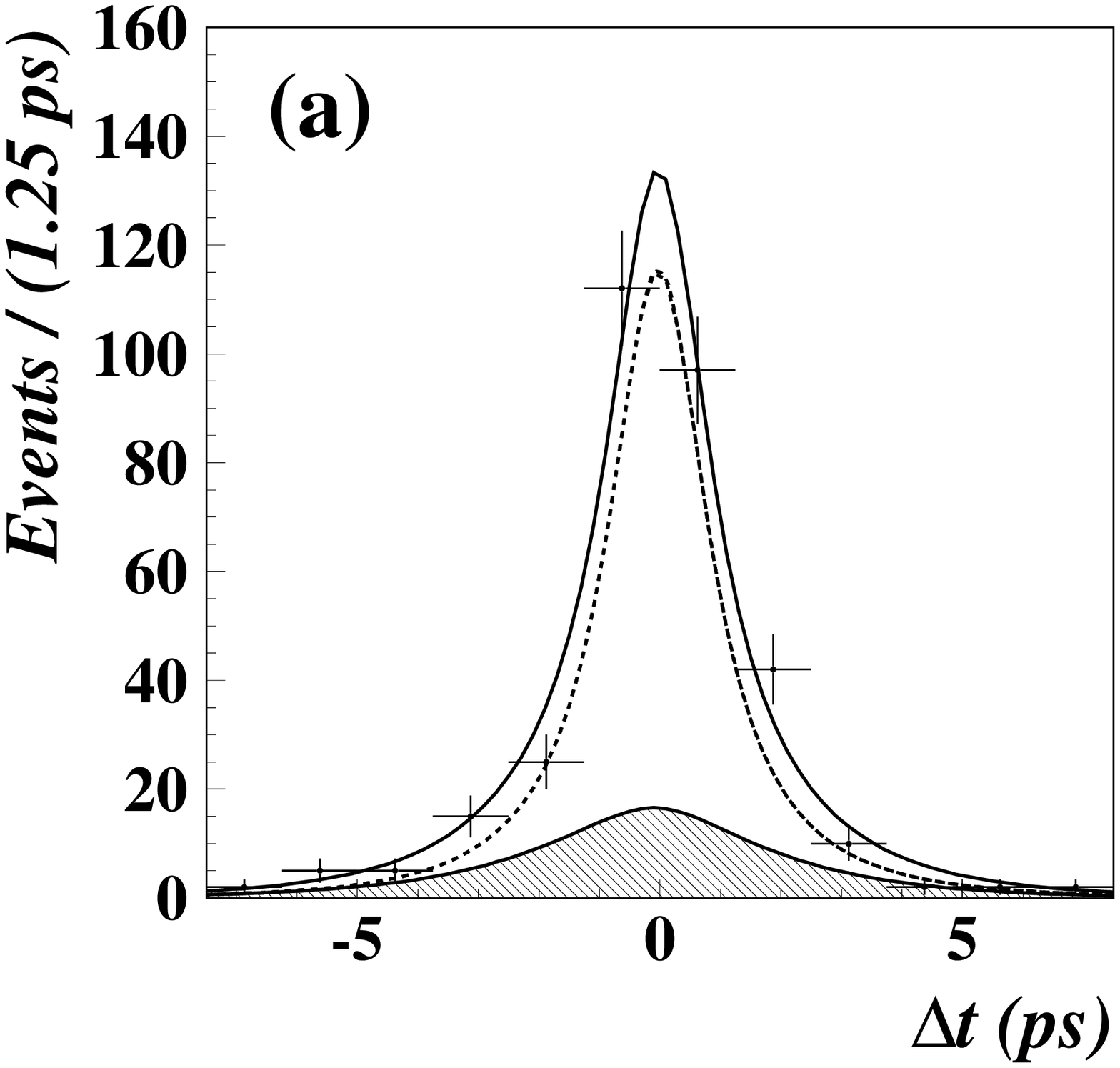}
\includegraphics[width=40mm]{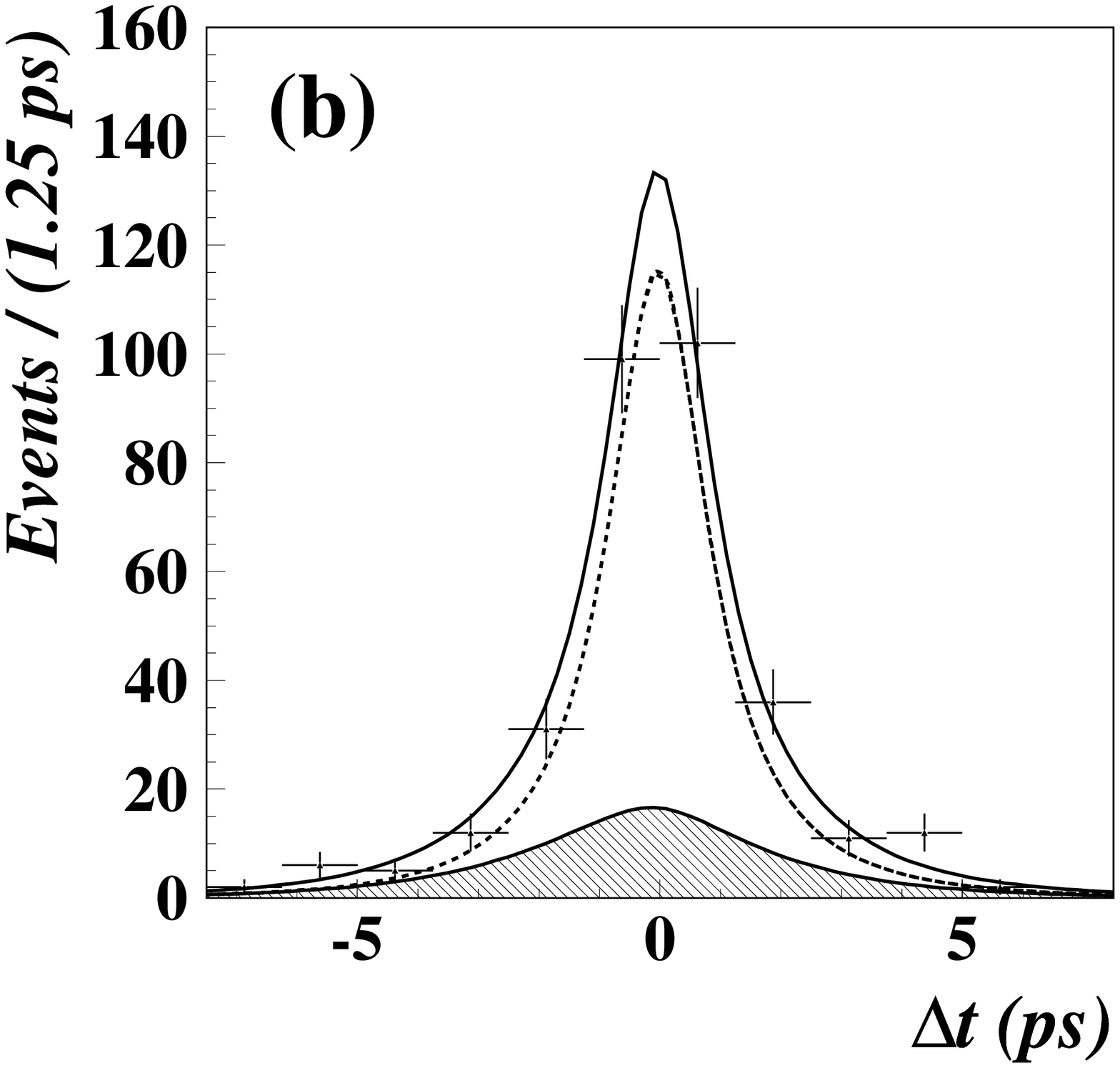}
\includegraphics[width=60mm]{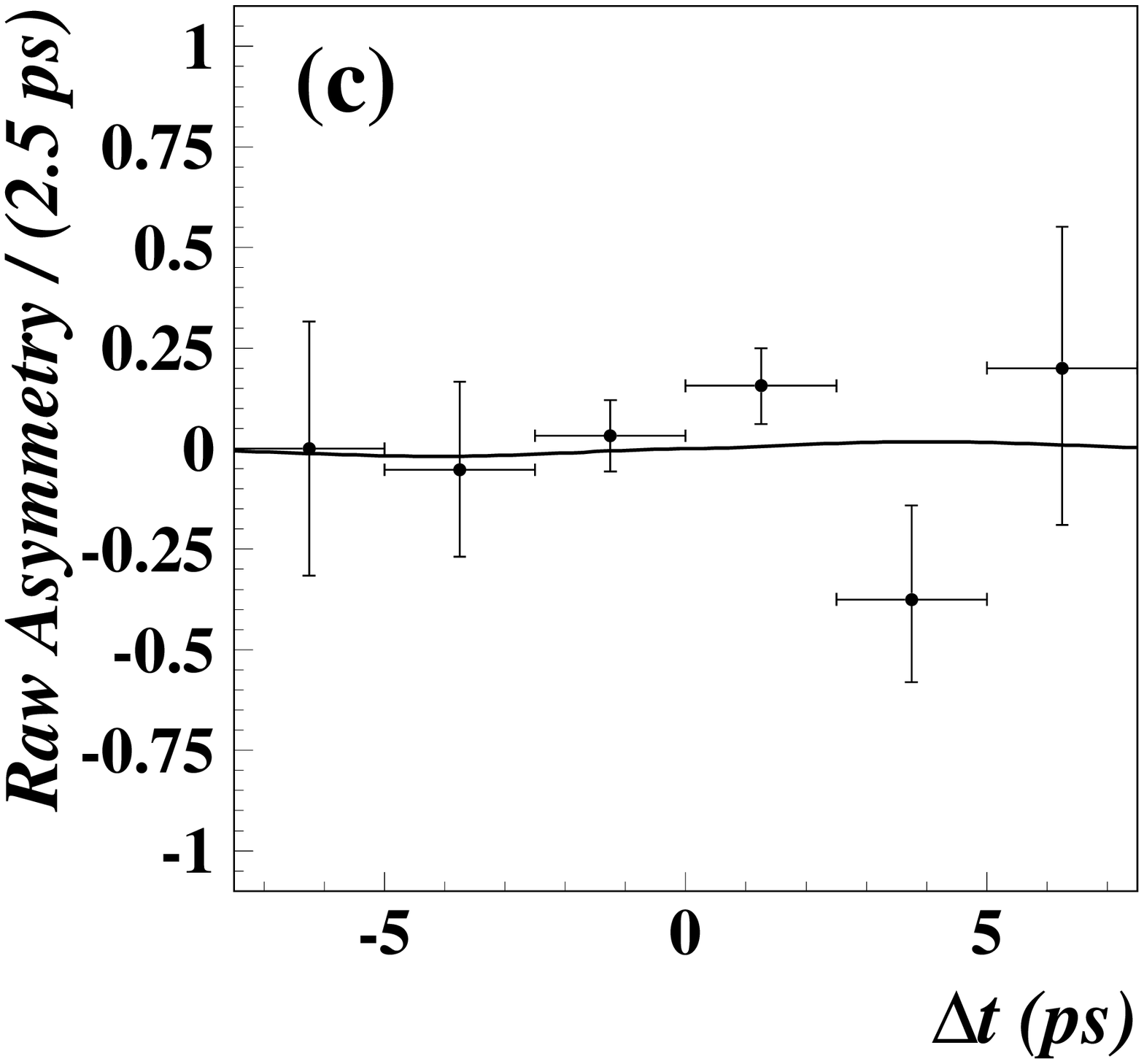}
\caption{The $\Delta t$ distribution inside $M_{bc}-\Delta E$ 
signal region and satisfy $0.62\rm{GeV/C^2}<M_{\pi\pi^0}<0.92\rm{GeV/C^2}$.
(a) and (b) are the $B^0$ tagged and $\overline{B^0}$ tagged events. (c)
is raw asymmetry for the events satisfy 0.5$<$r$\le$1.0. } 
\label{rhorho3_belle}
\end{figure}

The time-dependent $CP$ analysis from Belle is based on 
253~fb$^{-1}$~\cite{bib:belle_rhorho}. 
First, the longitudinal component 
fraction , $f_L = 0.941^{+0.034}_{-0.040}\rm{(stat)}\pm0.030\rm{(syst)}$, and
time-dependent parameters, 
$\mathcal{A}_L=0.00\pm0.30\rm{(stat)}\pm0.09\rm{(syst)}$  
and $\mathcal{S}_L=0.08\pm0.41\rm{(stat)}\pm0.09\rm{(syst)}$, are obtained. 
Figure~\ref{rhorho1_belle}, Figure~\ref{rhorho2_belle} and
Figure~\ref{rhorho3_belle} show the result of Belle's analysis.
Combining other related results~\cite{bib:babar_rho0rho0,bib:babar_rhorho_iso2,bib:belle_rhorho_iso2},
Belle obtain $59^\circ<\phi_2(\alpha)<115^\circ$ with 90\% C.L.

\begin{figure}[h]
\centering
\includegraphics[width=70mm]{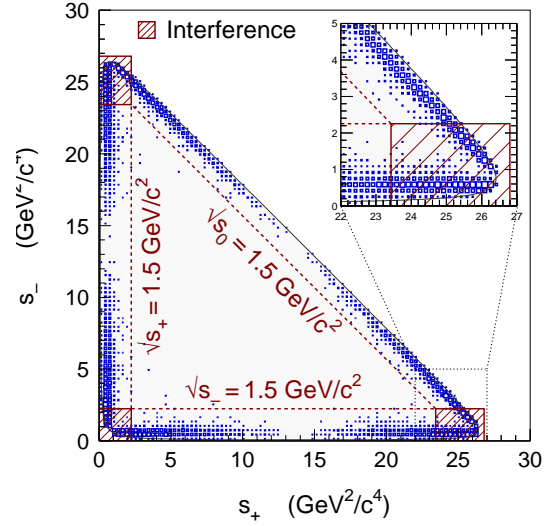}
\caption{The Dalitz plot from $B^0\to\pi^+\pi^-\pi^0$ Monte Carlo without detector simulation.
$\rho^+\pi^-$, $\rho^-\pi^+$ and $\rho^0\pi^0$ are generated with equal amplitudes.}
\label{rhopi1_babar}
\end{figure}

\begin{figure}[h]
\centering
\includegraphics[width=40mm]{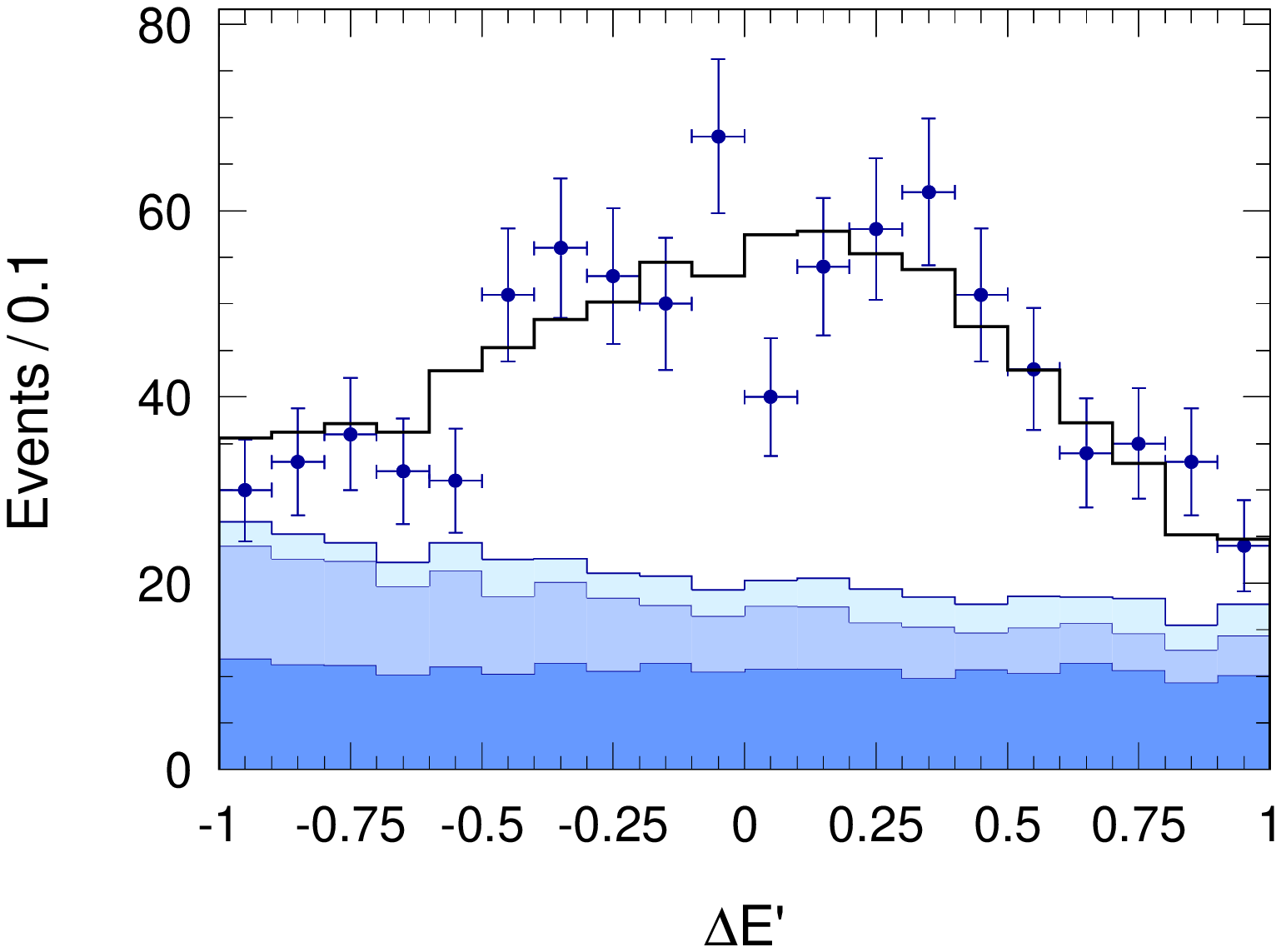}
\includegraphics[width=40mm]{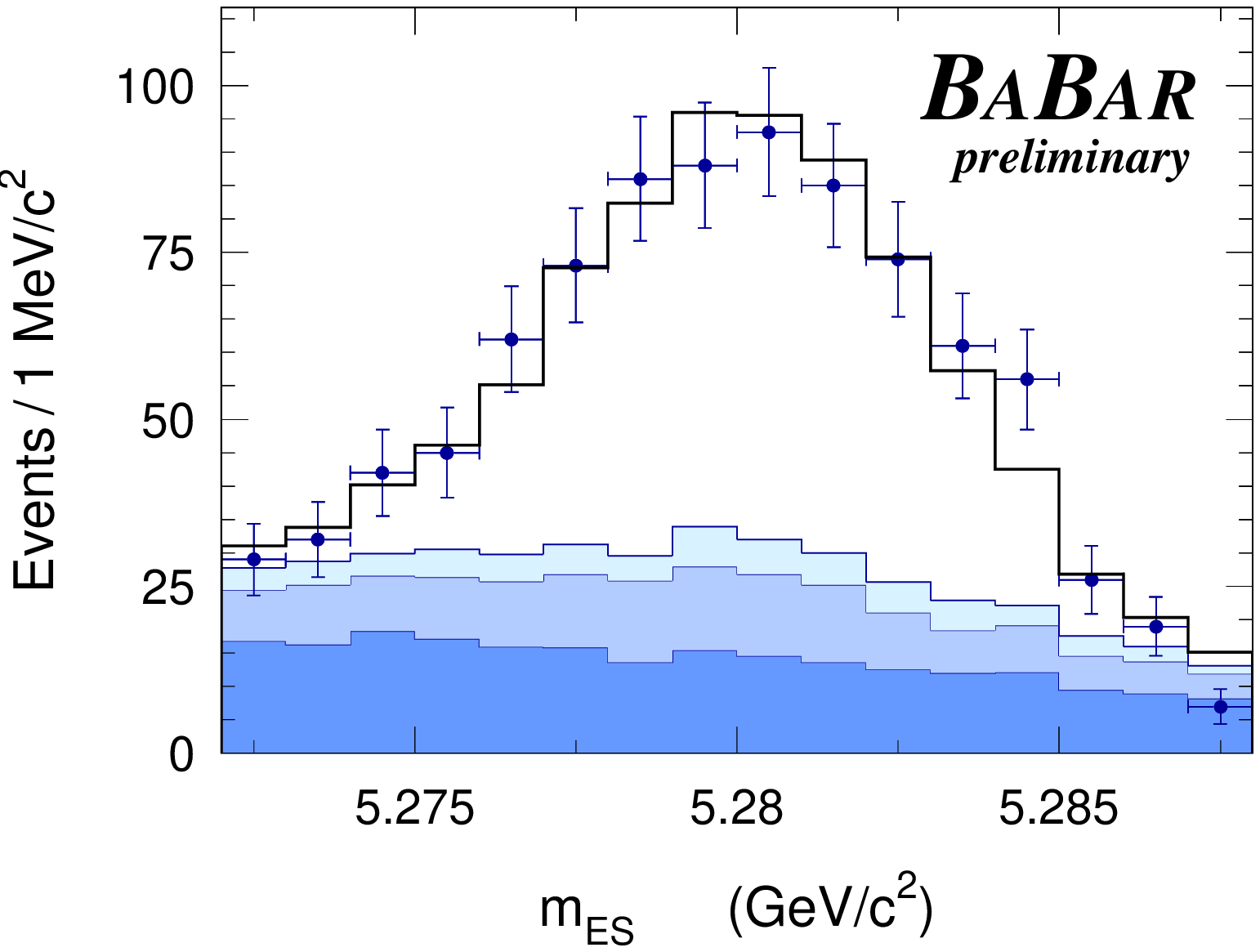}
\includegraphics[width=40mm]{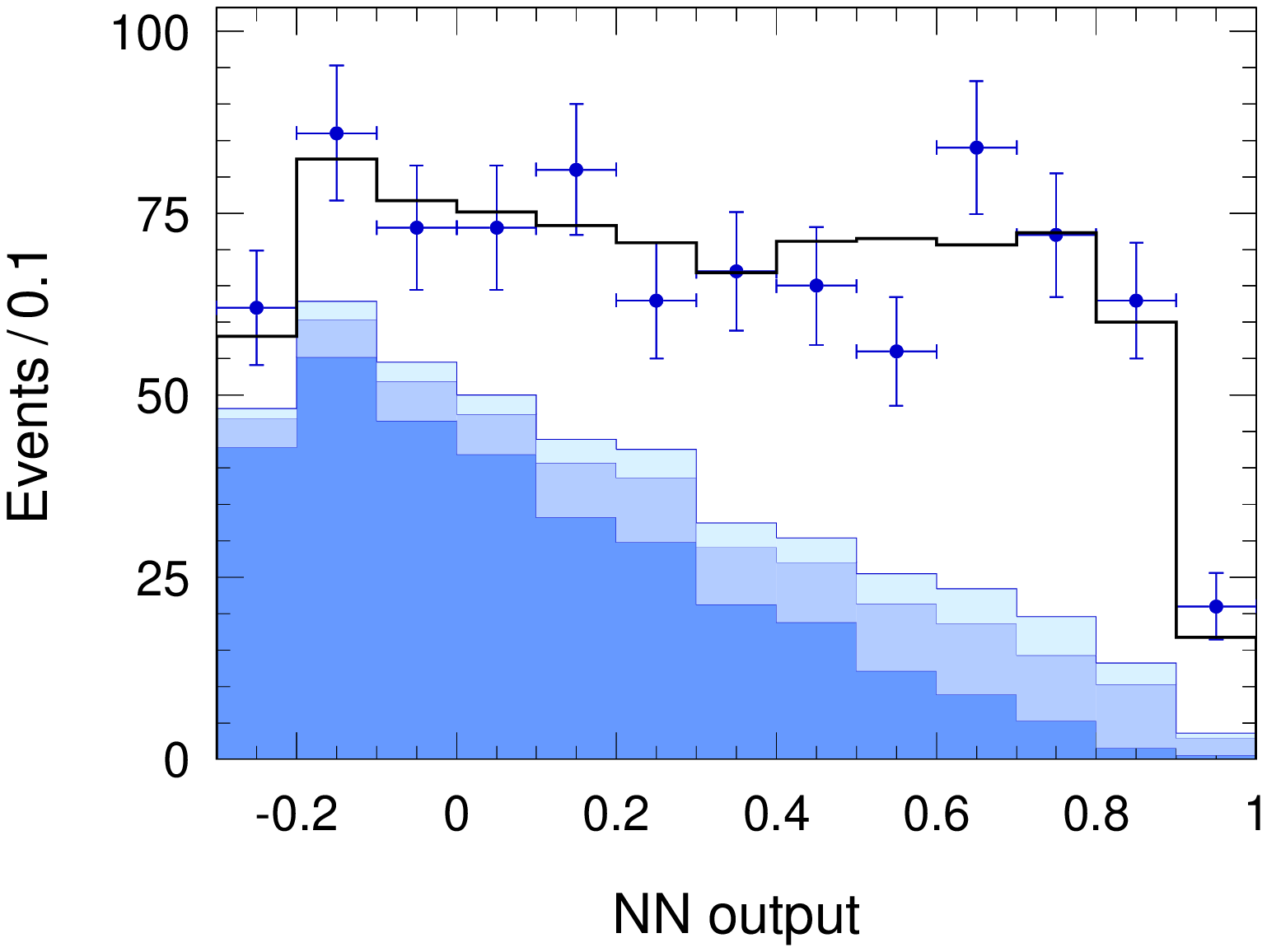}
\includegraphics[width=40mm]{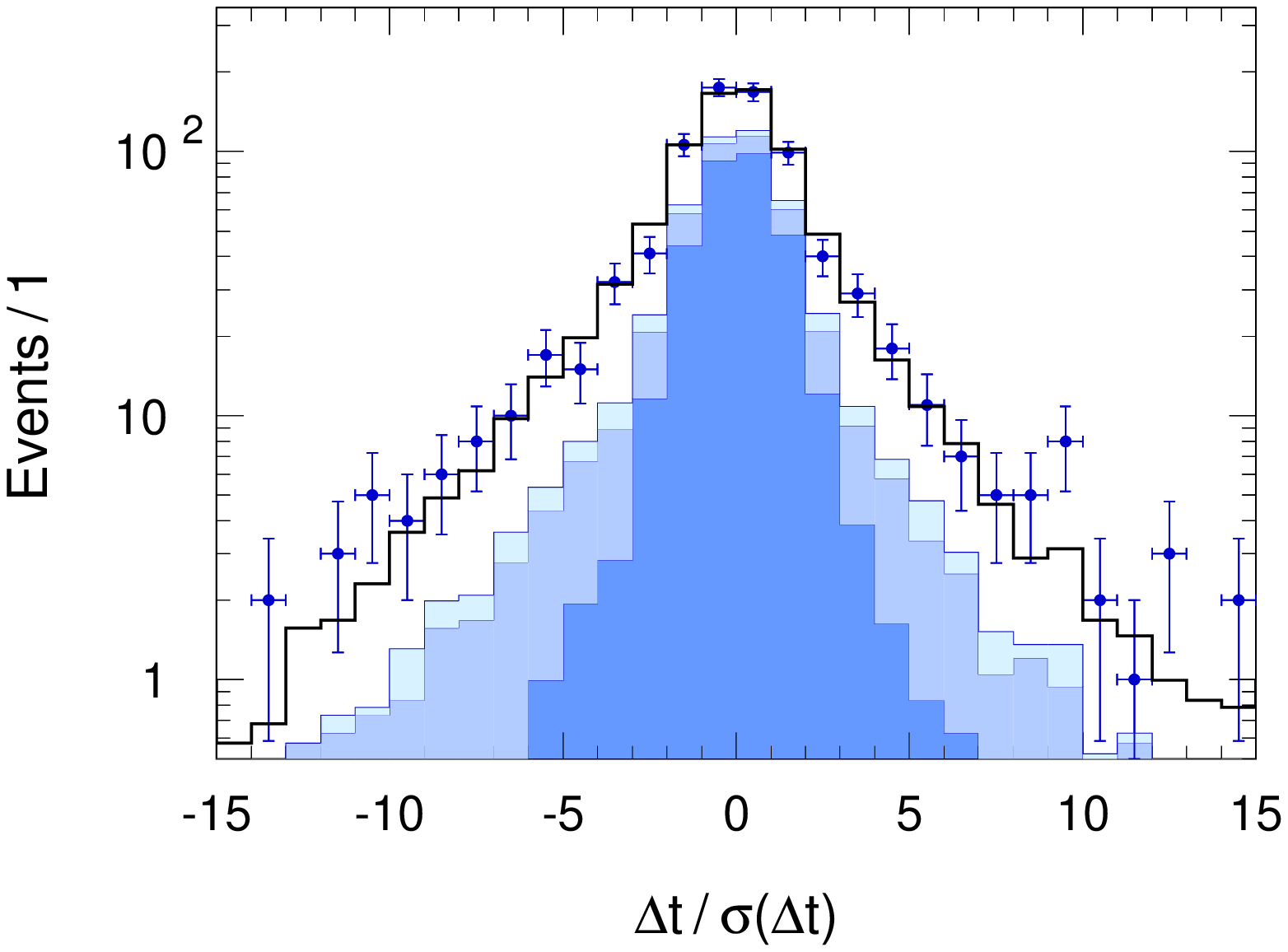}
\includegraphics[width=40mm]{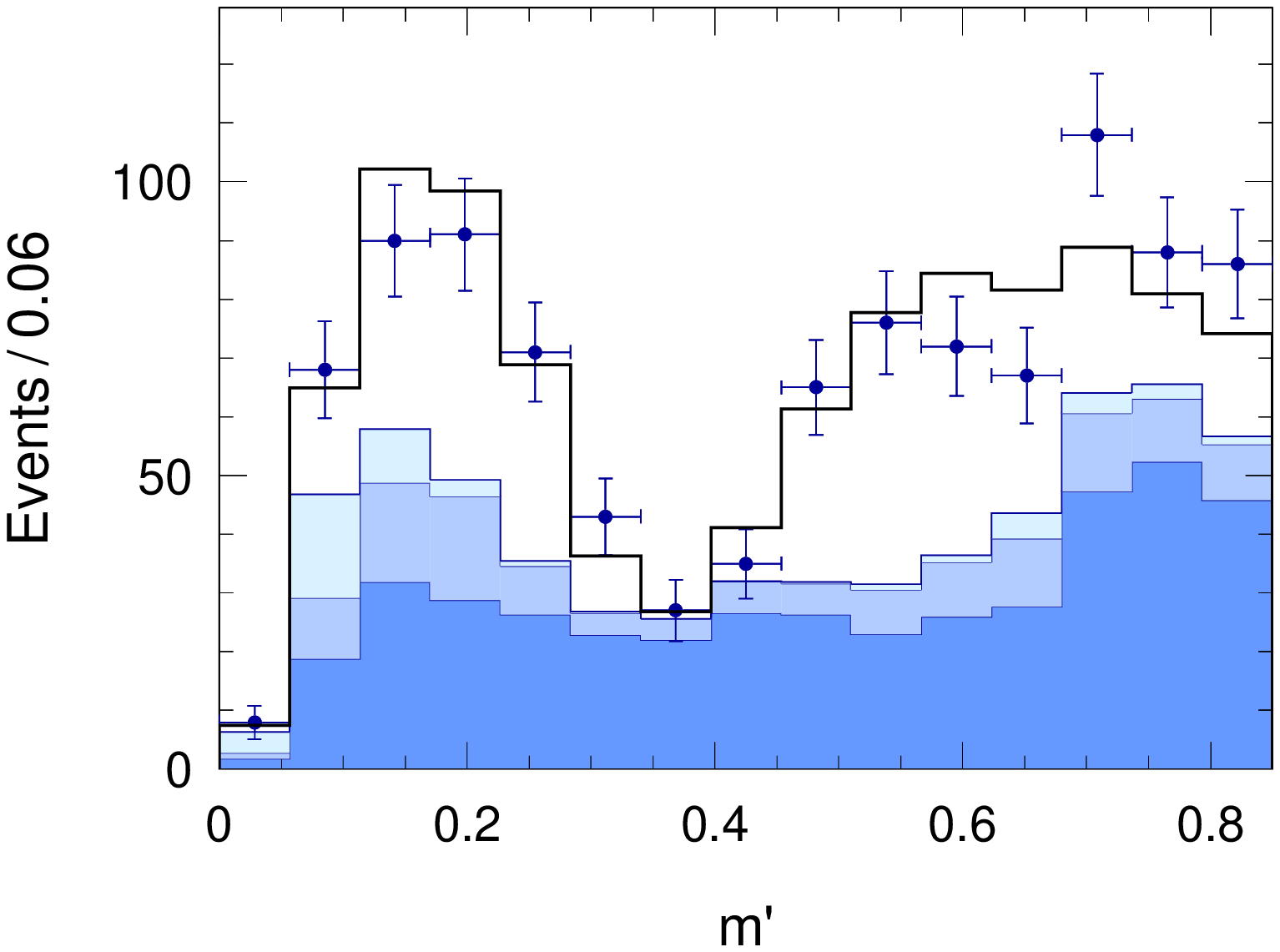}
\includegraphics[width=40mm]{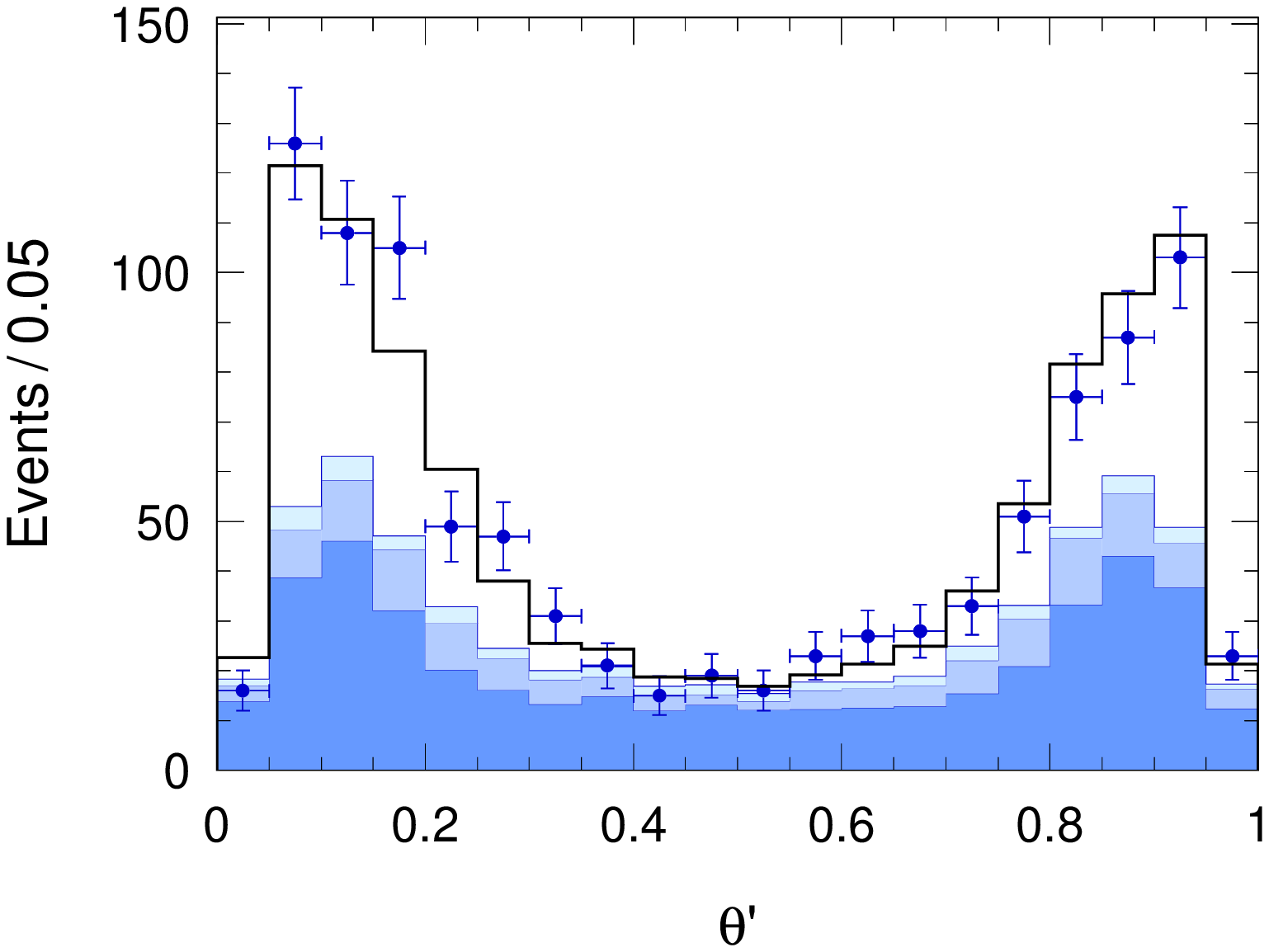}
\caption{Distribution of $\Delta E'$, $m_{ES}$, neural network output, 
$\Delta t/\sigma(\Delta t)$, m' and $\theta '$.  The dark, medium and 
light shaded areas represent the contribution from continuum events, the sum of continuum
events and the $B$ background, and the mis-reconstructed signal events, respectively.} 
\label{rhopi2_babar}
\end{figure}

\begin{figure}[h]
\centering
\includegraphics[width=70mm]{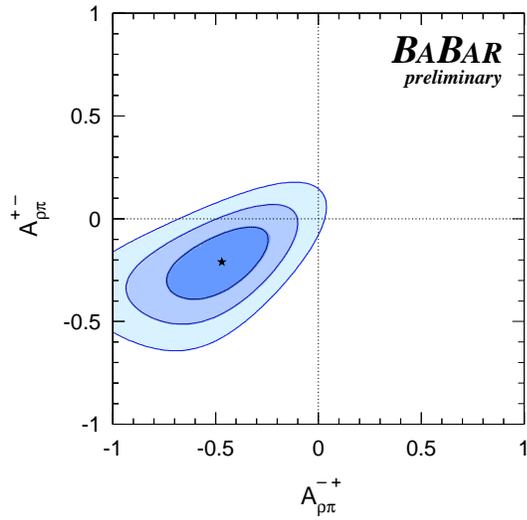}
\caption{Confidence level contours for the direct $CP$ violation. The shaded areas
represent 1$\sigma$, 2$\sigma$ and 3$\sigma$ contours, respectively. The $A^{+-}_{\rho\pi}$ and
$A^{-+}_{\rho\pi}$ here correspond to the Belle's $A^{-+}$ and $A^{+-}$, respectively.} 
\label{rhopi3_babar}
\end{figure}

\begin{figure}[h]
\centering
\includegraphics[width=70mm]{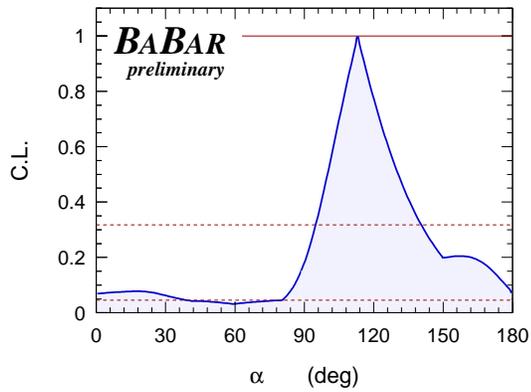}
\caption{Confidence level functions for $\phi_2/\alpha$. The dashed horizontal
lines corresponds to 1$\sigma$ and 2$\sigma$ C.L.} 
\label{rhopi4_babar}
\end{figure}

\section{$B\to\rho\pi$}
Besides $B\to\pi\pi$ and $B\to\rho\rho$ decays, the quasi-two-body analysis for
$B\to\rho\pi$ decay
is another candidate for the $\phi_2/\alpha$ extraction. Belle collaboration
performed the time-dependent analysis on $B\to \rho^\pm\pi^\mp$ with 140~fb$^{-1}$ 
and obtained time-dependent $CP$ parameters, 
$C_{\rho\pi} = 0.25^{+0.16}_{-0.17}\rm{(stat)} ^{+0.02}_{-0.06}\rm{(syst)},$
$S_{\rho\pi} = -0.28 ^{+0.23}_{-0.22}\rm{(stat)} ^{+0.10}_{-0.08}\rm{(syst)},$
$\Delta C_{\rho\pi} = 0.38 ^{+0.17}_{-0.18}\rm{(stat)} ^{+0.02}_{-0.04}\rm{(syst)}$ and
$\Delta S_{\rho\pi} = -0.30^{+0.24}_{-0.23}\rm{(stat)} \pm0.09\rm{(syst)},$
flavor integrated charge asymmetry,
$A^{\rho\pi}_{CP} =  -0.16^{+0.09}_{-0.10}\rm{(stat)} \pm 0.02\rm{(syst)}$,
and direct $CP$ violation parameters, 
$A_{+-} = -0.02^{+0.16}_{-0.15}\rm{(stat)} ^{+0.05}_{-0.02}\rm{(syst)}$ 
and $A_{-+} = -0.53 ^{+0.29}_{-0.28}\rm{(stat)} ^{+0.09}_{-0.04}\rm{(syst)}$~\cite{bib:belle_rhopi}. 

The branching fraction of $B^0\to\rho^0\pi^0$ is obtained to be
$3.12^{+0.88}_{-0.82}\rm{(stat)}\pm 0.33\rm{(syst)}^{+0.50}_{-0.68}\rm{(model)}\times 10^{-6}$,
and the $CP$ asymmetry, $A_{CP}=-0.53^{+0.67}_{-0.84}\rm{(stat)}\pm ^{+0.10}_{-0.15}\rm{(syst)}$, 
are obtained by Belle with 357~fb$^{-1}$ data~\cite{bib:belle_rho0pi0}.

Besides the quasi-two-body analysis, Babar performs the time-dependent 
Dalitz analysis from $B\to\rho\pi$ with $\pi^+\pi^-\pi^0$ final states~\cite{bib:babar_rhopi}
with 192~fb$^{-1}$ data. 
It is the first direct measurement of $\phi_2/\alpha$ by assuming isospin symmetry.  
The decay amplitude can be expressed as
$$A_{3\pi} = f_+A^++f_-A^- + f_0A^0,$$
$$\overline A_{3\pi} = f_+\overline A^++f_-\overline A^- + f_0\overline A^0.$$
The $f_+$, $f_-$ and $f_0$ are functions of $\pi^\pm\pi^0$ invariant mass that
incorporate the kinematic and dynamical properties of the $B^0$ decay into $\rho\pi$.
The Dalitz plot distribution for Monte Carlo is shown in Figure~\ref{rhopi1_babar}. 
The time-dependent rate for $B^0$ and $\overline{B}^0$ tagged events 
are $|A_{3\pi}^+(\Delta t)|^2$ and $|A_{3\pi}^-(\Delta t)|^2$, respectively. The rates are 
given by 
\begin{eqnarray}
\nonumber
|A_{3\pi}^\pm(\Delta t)|^2 & = & {e^{-|\Delta t|/\tau_{B^0}}\over 4\tau_{B^0}}
[|A_{3\pi}|^2 + |\overline{A}_{3\pi}|^2\\
\nonumber
& & \mp(|A_{3\pi}|^2 - |\overline{A}_{3\pi}|^2)\rm{cos}(\Delta m_d\Delta t)\\
\nonumber
& & \pm 2Im(\overline{A}_{3\pi}A^*_{3\pi}) \rm{sin}(\Delta m_d\Delta t)].
\end{eqnarray}

  There are 27 coefficients for this time dependent Dalitz rate, 9 for exponential,
9 for cosine oscillation and 9 for sine oscillation terms. Since the 
$B\to \rho^0\pi^0$ branching ratio is very small, the $B\to \rho^0\pi^0$ related oscillation
parameters are fixed to be zero and the effect is taken into account by systematics.
Figure~\ref{rhopi2_babar} shows the projection plots for $\Delta E'$, $m_{ES}$, 
neural network output for continuum background suppression, 
$\Delta t/\sigma(\Delta t)$, m' and $\theta '$. The $\Delta E'$ 
is the transformed $\Delta E$ to deal with the $\pi^0$ energy dependence and the  
$\sigma(\Delta t)$ is the event-by-event error on $\Delta t$. The m' and $\theta '$
are transformed Dalitz variables.
The fit yields the direct $CP$ violation values
$A^{+-}_{\rho\pi}= -0.21\pm0.11\rm{(stat)} \pm 0.04\rm{(syst)}$ and
$A^{+-}_{\rho\pi}= -0.47^{+0.14}_{-0.15}\rm{(stat)}\pm 0.04\rm{(syst)}$. 
Figure~\ref{rhopi3_babar} shows the confidence level of direct $CP$ violation. 
$\phi_2/\alpha$ obtained from this analysis is 
$(113^{+27}_{-17}\rm{(stat)}\pm 6\rm{(syst)})^\circ$ and confidence level
plot is shown in Figure~\ref{rhopi4_babar}.

\section{Summary}
\begin{figure}[h]
\centering
\includegraphics[width=70mm]{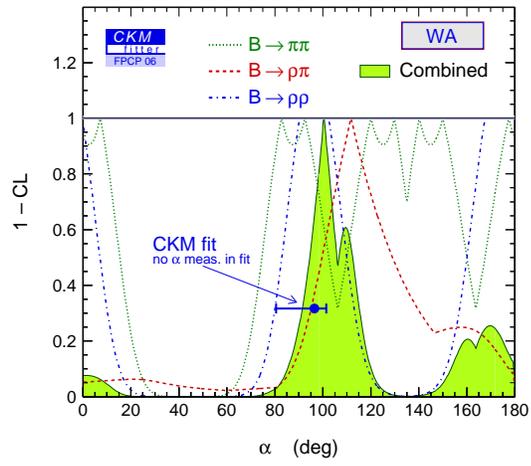}
\caption{The combined $\phi_2/\alpha$ constraint from CKM fitter Group.} 
\label{ckmfitter}
\end{figure}

From the $B\to\pi\pi$, $B\to\rho\rho$ and $B\to\rho\pi$ analysis, Belle
and Babar obtained $\phi_2/\alpha$ for each decay separately.
The CKM Fitter Group performs a global fit by properly averaging all the 
results and get $\phi_2/\alpha=(100.2^{+15.0}_{-8.8})^\circ$~\cite{bib:ckmfitter}.
Figure~\ref{ckmfitter} shows the result of this global fit for $\phi_2/\alpha$.


\bigskip 

\end{document}